\documentclass{rspublic}
\usepackage{psfig}
\usepackage{graphicx}

\def\bsax{{\it Beppo}SAX}
\def\sw{{\it Swift}}
\def\ama{$E_{\rm p}-E_{\rm iso}$}
\def\ghi{$E_{\rm p}-E_{\gamma}$}
\def\fir{$L_{\rm iso}-E_{\rm p}-T_{0.45}$}
\def\lia{$E_{\rm p}-E_{\rm iso}-t_{\rm break}$}
\def\nfn{$\nu F_{\nu}$}
\def\chisq{$\chi^2$}

\begin{document}

\title[GRB spectral correlations and cosmology]{GRBs spectral correlations and their cosmological use}

\author[G. Ghirlanda]{Giancarlo Ghirlanda}

\affiliation{INAF--Osservatorio Astronomico di Brera, Via E. Bianchi 46, I--23807,
  Merate (Italy)}

\label{firstpage}

\maketitle

\begin{abstract}{Gamma Rays: bursts, Cosmology}

The correlations involving the long--GRB prompt emission energy represent a
new key to understand the GRB physics. These correlations have been
proved to be the tool which makes long--GRBs a new class of standard
candles. Gamma Ray Bursts, being very powerful cosmological sources detected in
the hard X-ray band, represent a new tool to investigate the Universe
in a redshift range which is complementary to that covered by other
cosmological probes (SNIa and CMB). A review of the \ama\ , \ghi\ ,
\lia\ , \fir\ correlations is presented. Open issues related to these
correlations (e.g. presence of outliers and selection effects) and to
their use for cosmographic purposes (e.g. dependence on model
assumptions) are discussed. Finally, the relevance of thermal
components in GRB spectra is discussed in the light of some of the
models recently proposed for the interpretation of the spectral-energy
correlations.
\end{abstract}

\section{The \ama\ and \ghi\ correlations}

\begin{center}
\begin{figure}
\psfig{file=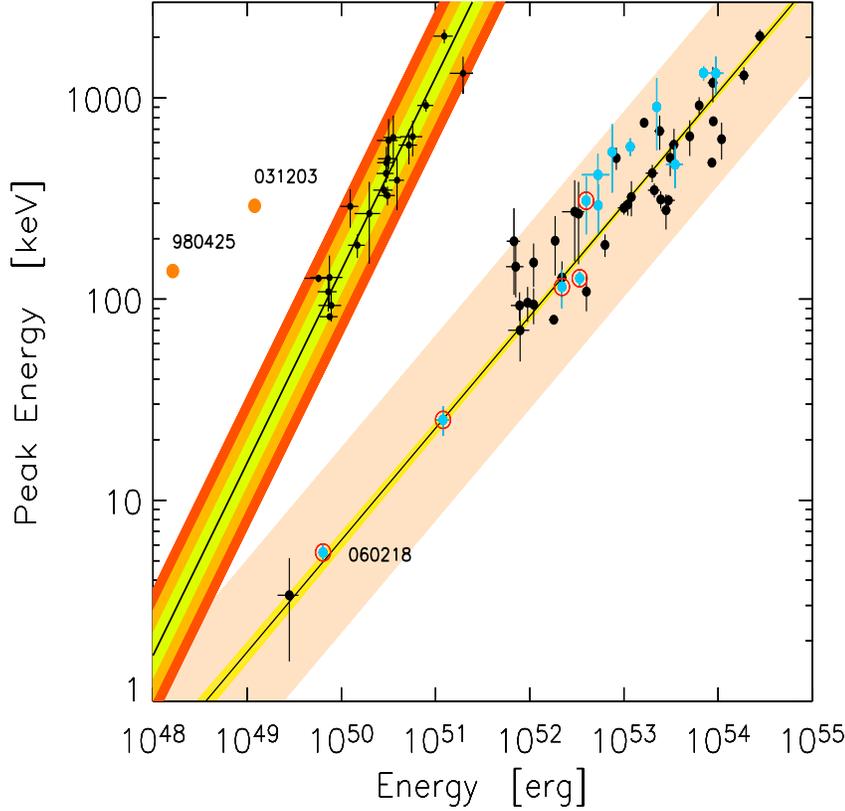,width=12cm,height=12cm}
\caption{Correlation between the \nfn\ peak spectral energy and the
isotropic energy (on the left side of the plot) defined with 49
GRBs (updated to 15 Sept. 2006). The best fit solid line is 
$E_{\rm p}/100 {\rm keV} = 3.9\pm 0.1 (E_{\rm iso}/1.62\times 10^{53})^{0.56\pm0.01}$
(\chisq = 530/47 dof) and the pale--orange
region represents the 3$\sigma$ scatter (computed perpendicular to
the best fitting line and modelled as a gaussian) of the data
points around the correlation. The blue points represent the 15
GRBs added since 2005 (i.e. in the \sw\ ``era''). Among these only
5 bursts had their $E_{\rm p}$ measured by Swift (red--circled
symbols). The two {\it outliers} with respect to the \ama\
correlation (GRB~980425 and GRB~031203) are shown. On the left
side of the plot it is shown the \ghi\ correlation (computed
in the case of a circumburst medium density scaling $\propto r^{-2}$, 
i.e. wind medium) with the most u1pdated sample of 21
GRBs. The solid line is the best fit to this correlation 
$E_{\rm p}/{100 {\rm keV}} = 3.6\pm 0.2 (E_{\gamma}/2.69\times10^{50})^{0.95\pm0.06}$
(\chisq = 17.4/19 dof) and the colored
regions represent the 1,2,3$\sigma$ scatter of the data points
around it.}
\end{figure}
\end{center}

Long GRBs with spectroscopically measured redshifts show a strong
correlation between the total {\it isotropic} energy emitted during
the prompt phase ($E_{\rm iso}$) and the peak energy of their \nfn\
spectrum ($E_{\rm p}$) computed in the source rest frame. This
correlation, discovered by Amati et al. (2002) with 12 long--GRBs detected
by \bsax, was confirmed by adding 23 bursts detected by other
satellites (Ghirlanda et al. 2004) and extended to very low energies
with few X-Ray Flashes (Lamb et al. 2004).  Figure~1 shows the \ama\
correlation updated to Sept. 2006 with 49 long GRBs. Since its
discovery only two bursts (GRB~980425 and GRB~031203) appeared
inconsistent with this correlation. Since the launch of the \sw\
satellite in Nov. 2004 only 13 out of $\sim$45 bursts with measured
redshifts (within the sample of $\sim$170 events detected by \sw\ )
were added to the \ama\ correlation. This is mainly due to the narrow
energy range (15--150 keV) of the BAT instrument on--board
\sw. Indeed, only in 5 cases (shown in figure 1) the peak energy was
measured from the BAT spectrum.

However, the energy derived under the isotropic assumption is huge and
widely dispersed (between few 10$^{50}$ and 10$^{54}$ erg). If,
instead, GRBs are collimated within a jet, the estimated energies and
their dispersion are highly reduced, i.e. by a factor
$f=(1-cos\theta)\simeq 10^{-4}-10^{-2}$ with $\theta\sim 1-10$ deg
(Frail et al. 2001). The observable consequence of the jetted nature
of GRBs is an (achromatic) break in their afterglow light
curves. Assuming a radiative efficiency of the prompt phase (20\%) and
the density profile of the circum--burst medium, either homogeneous
(HM) or wind--like (WM, e.g. scaling as $\rho_{\rm ism}\propto r^{-2}$), 
it is possible to estimate the jet opening angle from the
measure of the afterglow break time $t_{\rm jet}$ (Sari 1999).

In both scenarios (HM or WM) the {\it collimation--corrected} energy
$E_{\gamma}$ is tightly correlated with the peak energy $E_{\rm peak}$
(Ghirlanda et al. 2004; Nava et al. 2006).  While the \ama\
correlation requires only the knowledge of the GRB prompt emission
spectrum and of its redshifts, the \ghi\ correlation requires also the
measure of the jet break time from the afterglow light curve. The
\ghi\ correlation (in the WM case) is represented in figure~1 with the
most updated sample of 21 GRBs with measured $t_{\rm jet}$.

The \ghi\ correlation in the WM case is linear. This implies that: 1)
it is invariant for transformation from the source rest frame to the
comoving frame (i.e. both $E_{\rm p}$ and $E_{\gamma}$ transform
$\propto \Gamma^{-1}$ if looking a uniform jet within its opening
angle); 2) the total number of photons, in different GRBs, is roughly
constant $\sim 4\times 10^{56}$.

Note that the scatter of the collimation corrected correlation is
dominated by the statistical errors on the two variables, differently
from the \ama\ correlation which has a larger dispersion. The small
scatter is what makes the \ghi\ correlation a distance indicator and
allows to use GRBs as standard candles (Ghirlanda et al. 2004a; Firmani
et al 2005; Ghirlanda et al. 2006; Ghirlanda, Ghisellini \& Firmani 2006a).

\section{Open issues}

While possible interpretations of the \ama\ and the \ghi\
correlations have been recently proposed, there are still several open
issues about these correlations and their cosmological use.

\begin{center}
   \begin{figure}
    \psfig{file=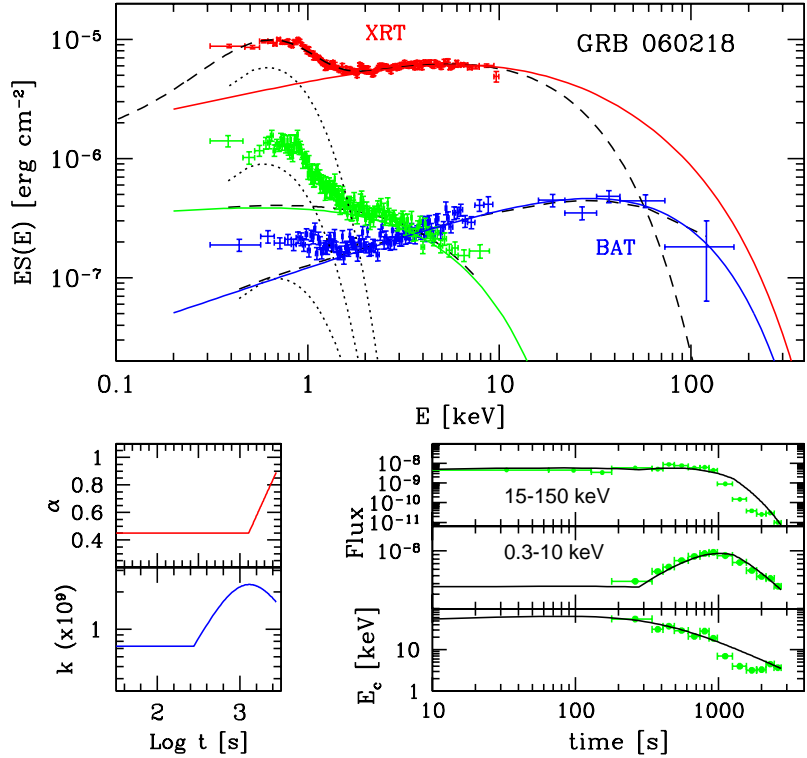,width=12cm,height=12cm}
   \caption{GRB~060218. {\it Top panel}: XRT and BAT spectra. Each
   spectrum is fitted with a cutoff--powerlaw plus a black body
   component (whose nature is discussed elsewhere, e.g. Campana et
   al. 2006; Ghisellini, Ghirlanda \& Tavecchio 2006a). Early and late
   time spectra (blue and green symbols) show a strong spectral
   evolution: the powerlaw spectral index and the \nfn\ peak energy
   soften, the latter moving from the 15--150 keV BAT energy range to
   the 0.3--10 keV XRT band. The red spectrum is integrated over the
   3000 s of duration of the burst and its cutoff--powerlaw model
   peaks at $E_{\rm p}\sim 5$ keV. The {\it right bottom panels} show
   the XRT and BAT light curves and the peak energy evolution. These
   are fitted self--consistently with the parameters evolving as shown
   in the {\it left bottom panels}.}
   \end{figure}
\end{center}

{\bf (1) Outliers:} GRB~980425 and GRB~031203 (both associated with a
nearby SN) are respectively five and four orders of magnitude
sub--luminous (but with a similar $E_{\rm p}$) with respect to the
population of bursts obeying the \ama\ correlation. It has been
proposed that they are normal GRBs observed off-axis
(e.g. Ramirez--Ruiz 2005). In this case, however, the true
luminosities of these two events (to be consistent with the \ama\
correlation) would make them the most luminous GRBs ever observed at
very low redshifts (i.e. 0.0085 and 0.106). This is hardly
reconcilable with any conceivable luminosity function. Instead, it
might still be the case that they are representative of a different
population of local sub--luminous GRBs (e.g. Soderberg et al. 2004).

An alternative explanation (Ghisellini et al. 2006), which aims at
testing if these two events can be consistent with the \ama\
correlation, was motivated by the recent \sw\ GRB~060218 (Campana et
al. 2006) also associated with a SN event at $z=0.033$. Its total
isotropic energy ($\sim 7\times 10^{49}$ erg) is only slightly larger
than that of the two outliers. Nonetheless, its very long duration
(3000 s) coupled with a strong hard--to--soft spectral evolution
(figure 2) makes its time-averaged spectral peak energy $\sim 5$ keV,
i.e. fully consistent with the \ama\ correlation.

Using GRB~060218 as a template we tried to model the spectral
evolution of GRB~031203 and 980425 with the available data.  It turns
out that in these two bursts a strong spectral evolution might have
caused part of their energy to be emitted in the soft X--ray band
where it went undetected. In the case of GRB~031203 (figure 3),
indeed, there is evidence that a late time soft X--ray fluence,
comparable to that observed in the $\gamma$--ray band, might be
responsible for the observed dust scattering halo evolution (Tiengo \&
Mereghetti 2006). As a result the total energy of these two events is
only slightly larger than what measured from their $\gamma$--ray
spectra while their peak energy is considerably (a factor 10--20)
smaller.

\begin{center}
   \begin{figure}[t]
     \psfig{file=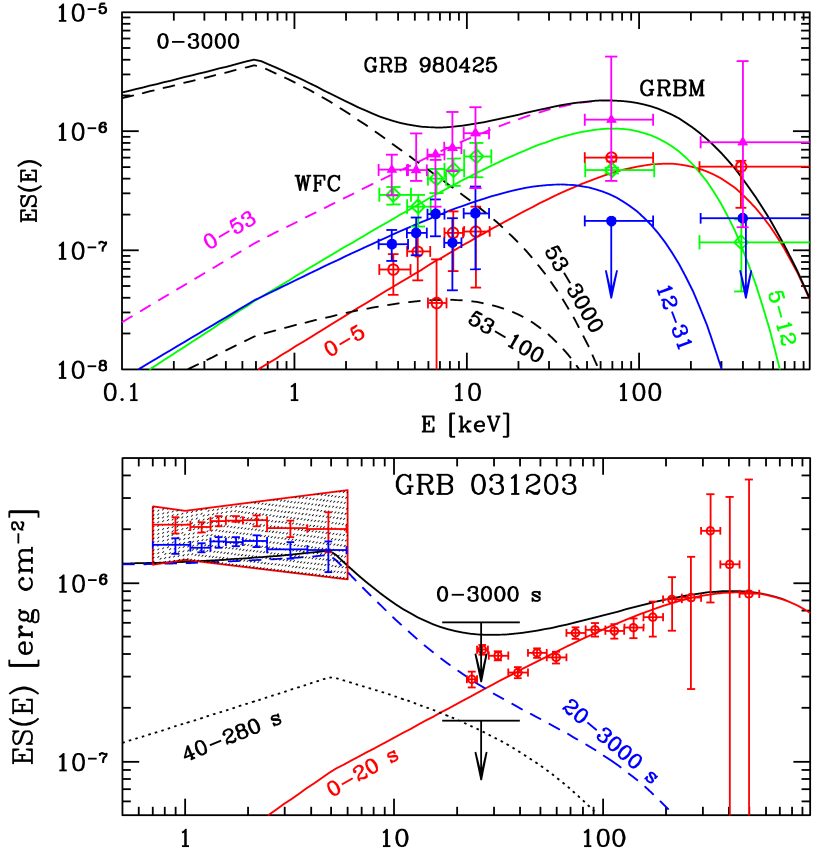,width=12cm,height=12cm}
   \caption{{\it Top panel}: spectral evolution of GRB~980425. The
   data are from \bsax (WFC: 2--28 keV and GRBM: 40--700 keV adapted
   from Frontera et al. 2000). The model fits (lines) are obtained
   with the same model used for GRB~060218 by simultaneously fitting
   the light curves and the available spectra of GRB~980425. {\it
   Bottom panel}: spectral evolution of GRB~031203. In this case the
   late time spectrum should produce a considerable flux in the X--ray
   band to be consistent with the observed evolution of its dust
   scattering halo (Tiengo \& Mereghetti 2006).}
   \end{figure}
\end{center}

{\bf (2) Selection effects:} It has been argued that different
samples of BATSE bursts, without a redshift measure, are inconsistent with
the \ama\ correlation for any distance they might be located at (Nakar
\& Piran 2005, Band \& Preece. 2005, Kaneko et al. 2006).

We note that: I) the updates of the \ama\ correlation (Ghirlanda
2004, Lamb 2004, Amati 2006 and the present paper - figure~1), i.e. from
12 to 49 events, show that all bursts with measured $z$ and
spectral properties do follow this relation (i.e. the outliers are still only
2 bursts); II) a test (Ghirlanda et al. 2005), performed with 442
GRBs for which only a pseudo redshift estimate is available (from
the Lag--Luminosity correlation - Norris, Marani \& Bonnel 2000), has confirmed
that these bursts still define a correlation in the \ama\ plane.
This correlation has a similar slope and a different normalization
(but only a slightly larger scatter) with respect to the
correlation defined with the GRBs with spectroscopically measured redshifts.
On the other hand, Nakar \& Piran 2005, Band \& Preece 2005 were unable to
argue that more than a small fraction of that only a small fraction of GRBs
are inconsistent with the \ghi\ correlation. Therefore, if we assume that the \ghi\ 
correlation is true we can derive from a given $E_{\gamma}$ its isotropic

equivalent. If the GRB angle distribution were uniform
the probability to derive any value of $E_{\rm iso}=E_{\gamma}/f$ would
be equal. This would produce a random (nearly) uniform scatter of
data points in the \ama\ plane. Instead, if the angle distribution
is peaked we should find a clustering of the data points around
some correlation. The comparison of the angle distribution of the 442
GRBs with pseudo redshifts is indeed peaked (figure 4). The
different average angle of the two distributions might be due to the
preference of detecting the most luminous GRBs, i.e. those with (on average)
a smaller jet opening angle.

\begin{center}
   \begin{figure}
\resizebox{13cm}{7cm}{\includegraphics{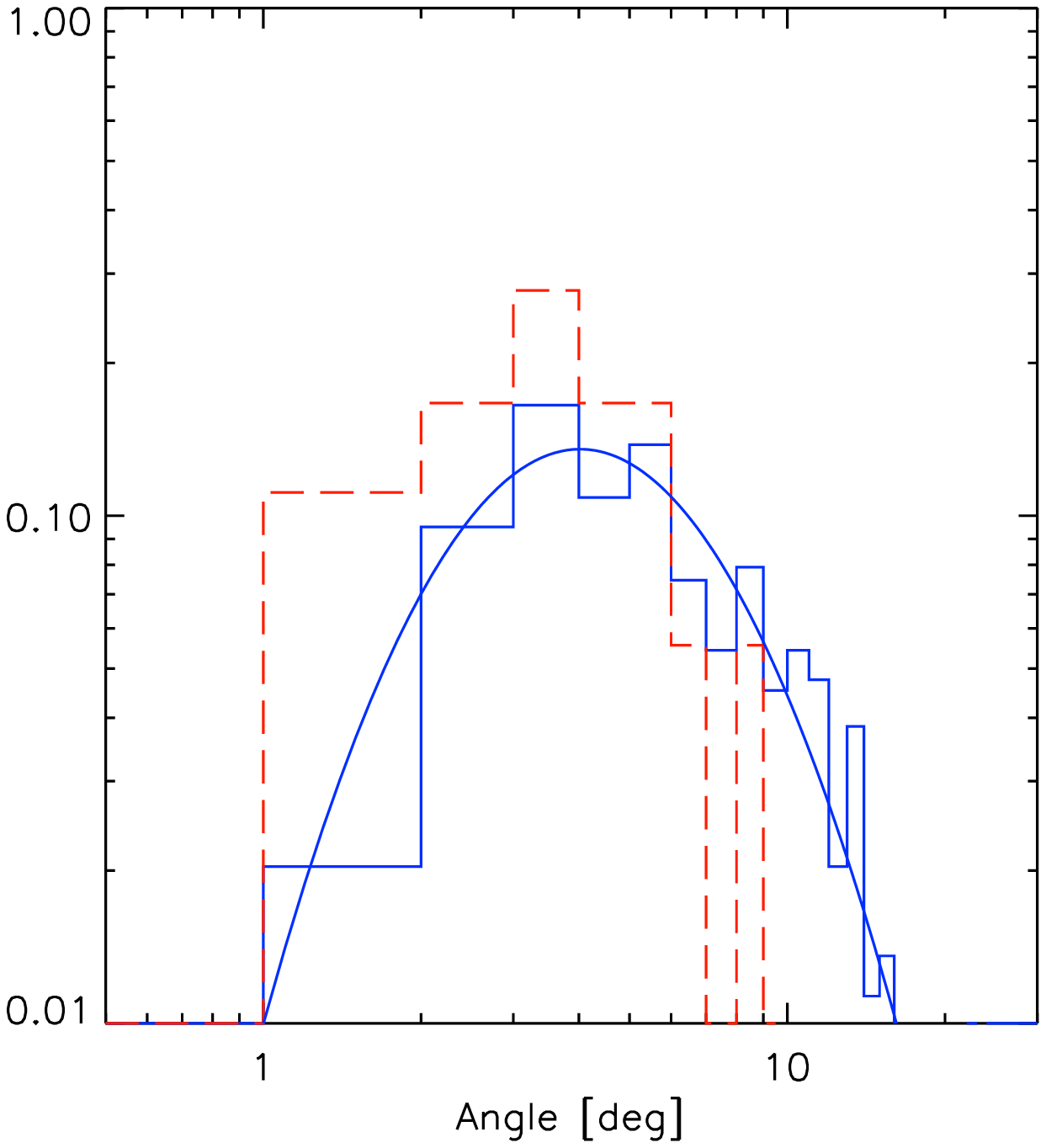}
\includegraphics{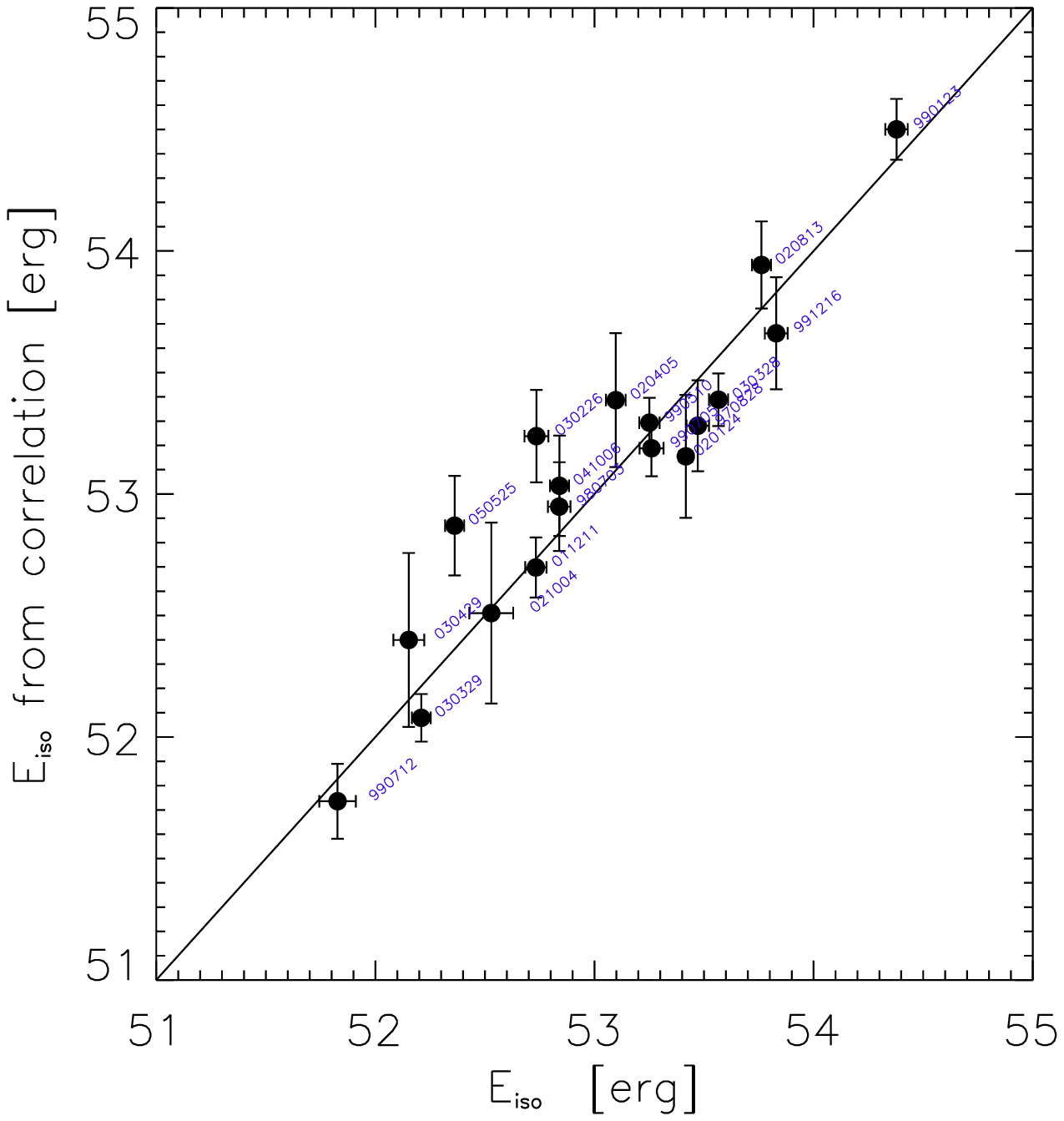}}
   \caption{{\it Left panel}: jet opening angle distribution of 21
   GRBs with measured $z$ and $t_{\rm break}$ (dashed line). The
   opening angle distribution of the large sample of 442 GRBs with
   pseudo-z measured from the Lag--Luminosity correlation (Band,
   Norris \& Bonnel 2004) is shown (solid line histogram) together
   with its log--normal fit (solid line - with 
$\langle \theta_{\rm jet}\rangle \sim 6$ deg). {\it Right panel}: the \lia\
   correlation. The best fit is 
$E_{\rm iso}=1.12\pm0.11 (E^{\prime}_{\rm p}/295 keV)^{1.93\pm0.17}\cdot (t^{\prime}_{\rm jet}/0.51 d)^{-1.08\pm0.17}$, 
where primed quantities are in the    source rest frame.}
   \end{figure}
\end{center}

{\bf (3) Model dependence of the \ghi\ correlation.} The \ghi\ correlation
is derived in the standard uniform jet scenario assuming a constant
radiative efficiency and a circumburst medium density
profile. Although the present afterglow observations do not allow to
distinguish between the homogeneous or the wind density circum--burst
scenario, the properties of the \ghi\ correlation (small scatter and
linear slope) derived in the WM case are appealing (Nava et al. 2006)
also for the improvement of the cosmological constraints (Ghirlanda et
al. 2006). However, the model dependence of this correlation still
represents one of its main weak points.  Liang \& Zhang (2005)
discovered a completely empirical correlation $E_{\rm iso}\propto E_{\rm p}^{2} t_{\rm break}^{-1}$ 
(figure 4 - Nava et al. 2006). Through this correlation it is possible to derive
cosmological constraints which are consistent with those obtained with
the two (HM and WM) model dependent \ghi\ correlations. It has been
demonstrated (Nava et al. 2006) that the \lia\ correlation is fully
consistent with the two model dependent correlations and this
strengthen the possibility to use GRBs as standard candles.

\begin{center}
   \begin{figure}
\resizebox{15cm}{8cm}{\includegraphics{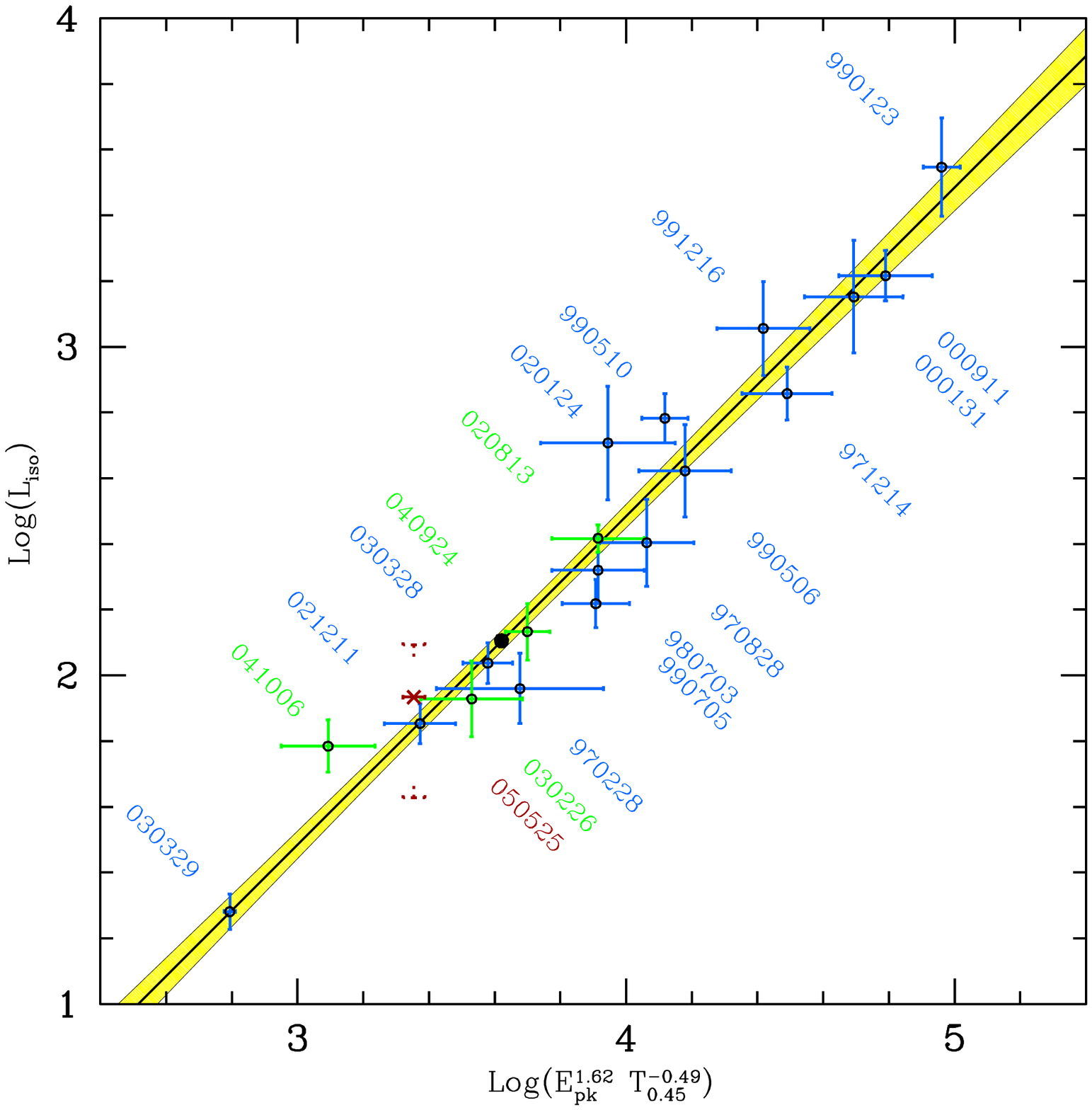}
\includegraphics{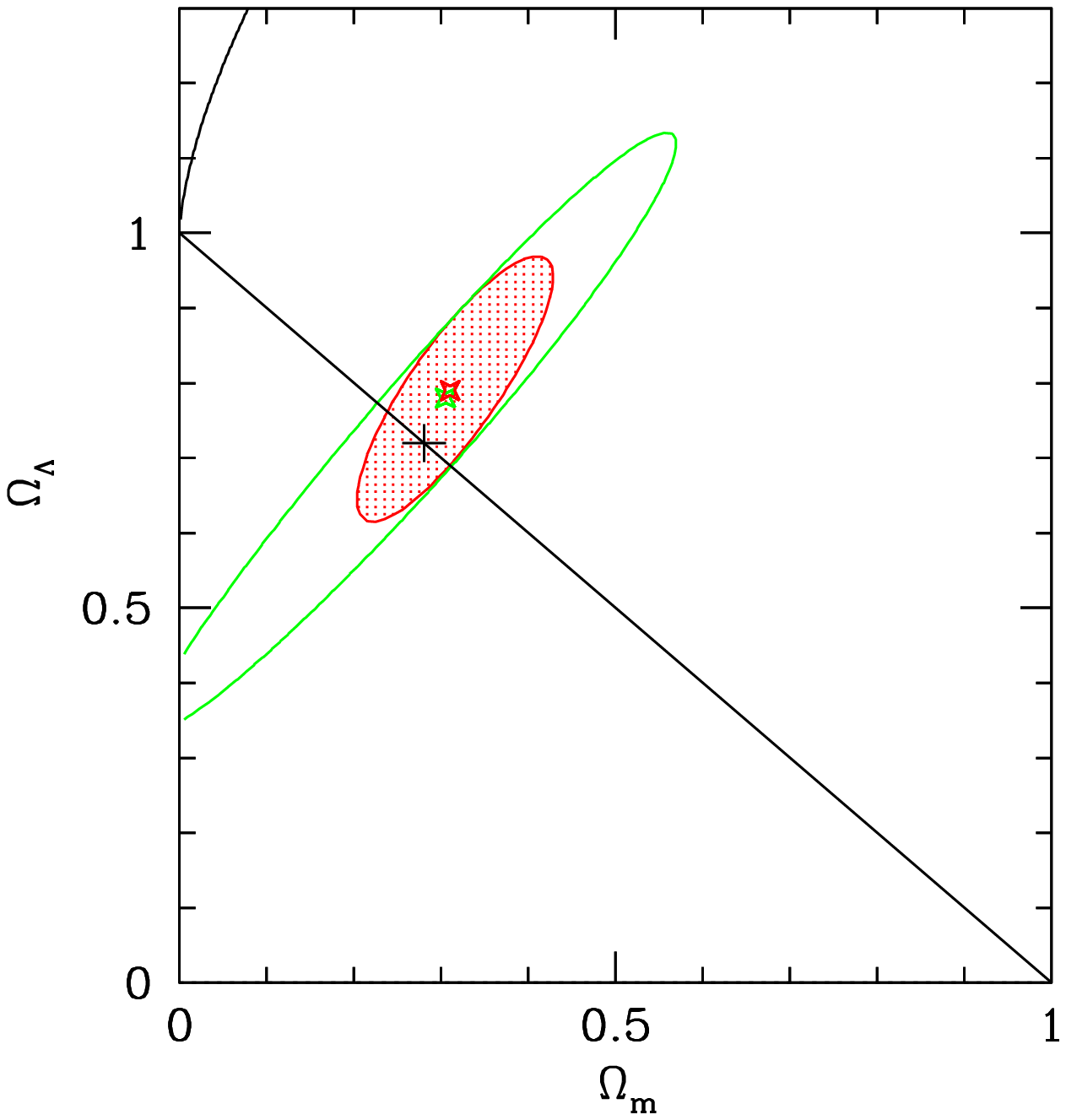}}
   \caption{{\it Left panel}: the \fir\ correlation defined with the
   most updated sample of GRBs with firmly measured spectral
   properties and light curves (from Firmani et al. 2006). The solid
   line is the best fit to this
   correlation, i.e. 
$L_{\rm iso}=10^{52.11\pm0.05}(E_{p}/234.4\ {\rm keV})^{1.62\pm0.08}\ (T_{0.45}/2.88\ {\rm s})^{-0.49\pm0.07}$. 
{\it Right panel}: the
  cosmological constraints on the $\Omega_{M}$-$\Omega_{\Lambda}$
   plane obtained with the Legacy SNIa sample (green line) and
   combining these with 19 GRBs (solid filled contours. Only the 68\%
   confidence contours are shown.}
   \end{figure}
\end{center}

One of the main still open issues related to the \ghi\ and \lia\
correlations is the fact that they require the measure of the
afterglow jet break time $t_{\rm break}$. While most of the jet breaks
of the ``gold'' sample of 21 bursts used to define these correlations
are derived from the optical afterglow light curves, there is growing
evidence that several \sw\ bursts do not show a break in their X-ray
and optical light curves when it should be expected according to these
correlations.

Although the lack of jet break times is still an open issue to be
investigated, the use of GRBs as standard candles has been definitely
confirmed by the recent discovery of a new correlation which is not
affected by this problems. Firmani et al. (2006) found that there is a
very tight correlation between the GRB isotropic luminosities $L_{\rm iso}$, 
the peak energy $E_{\rm p}$ and a characteristic timescale of
the prompt emission light curve $T_{0.45}$ (figure 5). The latter
parameter was originally defined to compute the GRB variability
(Reichart et al. 2001) which is indeed correlated with the GRB
luminosity. This new \fir\ correlation, being model independent and
assumptions--free, solves the previous problems. Moreover, the
cosmological constraints derived with this correlation, though still
based on a small sample of GRBs, are tighter than those obtained with
the \ghi\ correlations (Firmani et al. 2006a). The larger redshift
extension of GRBs with respect to SNIa and the fact that the \fir\
correlation is based only on prompt emission properties (i.e. related
to the detection of the GRB prompt emission in the $\gamma$--ray band)
makes GRBs a new cosmological tool complementary to SNIa. By adding 19
GRB to the sample of 115 SNIa (Astier et al. 2005) the constraints on
the $\Omega_{\rm M, \Lambda}$ (as well as on the parameters describing
the dark energy equation of state - Firmani et al. 2006a) are
considerably improved (figure 5) and show that GRBs and SNIa seem to
prefer the $\Lambda CDM$ model.

\begin{center}
   \begin{figure}
     \psfig{file=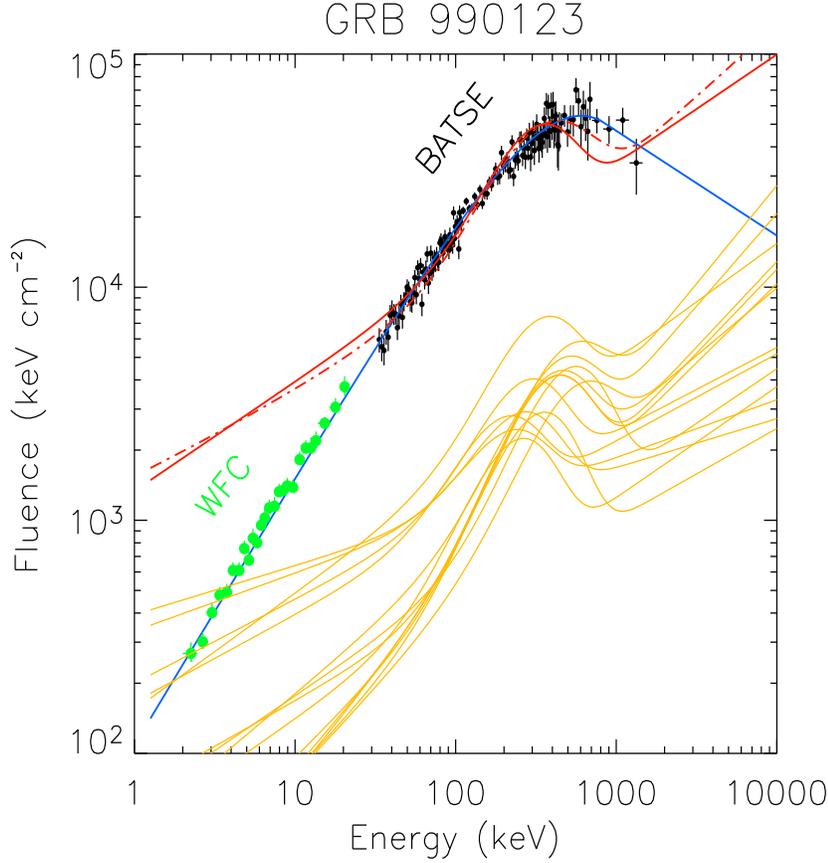,width=12cm,height=12cm}
   \caption{Prompt emission spectrum of GRB~990123 time--averaged over
     its duration: BATSE/CGRO data (black points) and WFC/\bsax\ data (green
     points). Solid blue line is the best fit with the non--thermal Band
     empirical model. The fit of the BATSE data alone with the mixed model
     (i.e. black body + powerlaw) is good (solid red line) but the resulting
     powerlaw component is inconsistent with the WFC data. Note also that the
     WFC and BATSE (low energy) data are consistent with a single powerlaw
     ($EF(E)\propto E^{1.15}$, which excludes the possibility that it is the
     self absorbed synchrotron emission) extending from 2 keV to few hundred
     keV. Solid orange lines are the model fits with the mixed black body +
     powerlaw model to the time resolved spectra covering the duration of the
     burst. Their sum is represented by the dot--dashed red line.}
   \end{figure}

\end{center}

{\bf (4) Thermal components in GRB spectra and the interpretation of
the \ama, \ghi, \fir\ correlations.} Among the proposed
interpretations of the above correlations (Levinson \& Eichler 2005,
Toma et al. 2005) Rees \& Meszaros (2005) suggested that a thermal
black body spectrum is the most natural way to link the peak spectral
energy and the total luminosity of GRBs as shown by the \ama\
correlation (see also Thompson 2006; Thompson, Meszaros \& Rees
2006). This ``thermal'' interpretation requires that the prompt
emission spectrum of GRBs is dominated by a thermal black body which
determines the peak of the \nfn\ spectrum. Thermal emission is
expected in the standard ``hot'' fireball model (Goodman 1986): it is
the initial black body which survived to the conversion (during the
opaque--acceleration phase) into bulk kinetic energy. An alternative
scenario (Rees \& Meszaros 2005) proposes that the thermal photons are
created by dissipation below the GRB photosphere.

Evidences of the presence of a thermal black body component were
discovered in the BATSE spectra (Ghirlanda, Celotti \& Ghisellini 2003; Ryde 2004)
although this component dominated the initial phase of $\sim 2$ sec of
the prompt emission. During this phase it was shown that the
luminosity and the temperature evolve similarly in different GRBs
while the late time spectrum is dominated by a non thermal component
(e.g. fitted with the empirical Band et al (1993) model). Attempts to
deconvolve these spectra with a mixed model, i.e. a thermal black body
and a non thermal powerlaw (Ryde et al. 2005), showed that the presence
of the black body component (with a monotonically decreasing
flux) could be extended to the late prompt emission phase (see also
Bosnjak et al. 2005).

In order to test the applicability of the thermal interpretation to
the \ama, \ghi, \fir\ correlations, we have been verifying if a
thermal component can be fitted to the spectra of the bursts that are
used to define these correlations (Ghirlanda et al. 2007). Ryde et
al. 2005 showed that the mixed model fits to the time resolved spectra
of GRBs is almost equivalent to a fit with a non thermal model ``a la
Band''. We therefore selected the 10 GRBs on the \ama correlation
detected by BATSE: these data allow to analyze the spectral evolution
with adequate spectral resolution. We succeeded in fitting a thermal
black body component plus a powerlaw. In general (as also found by
Ryde 2005 in few GRBs) the powerlaw component softens during the burst
being (on average) $F(E)\propto E^{-0.5}$ at the very beginning. The
black body also evolves in time and comprises at most $\sim$50\% of
the total spectral flux. However, such a soft non-thermal powerlaw
component should dominate the spectrum in the X--ray energy band. We
found that in 5/10 GRBs there are also X-ray data from the \bsax/WFC
(2-28 keV) and in all these cases the data of the WFC are inconsistent
with the extrapolation in the 2-28 keV band of the powerlaw fitted to
the BATSE $\gamma$--ray data. Moreover the X--ray to $\gamma$--ray
(WFC+BATSE) broad band spectrum is consistent with a single
non--thermal fit (with the Band model) with a low energy spectral
component much harder than the powerlaw of the mixed model. An example
is shown in figure 6. These results represent a challenge to the
presence of a dominating black body component in these bursts.


   \begin{acknowledgements}
   I am grateful to G. Ghisellini, C. Firmani, F. Tavecchio,
   A. Celotti, L. Nava, M. Nardini, Z. Bosnjak for years of fruitful
   collaboration. I am also grateful to M.L. \& T. Smith for the
   valuable logistic support during my stay in London.
   \end{acknowledgements}

\label{lastpage}
\end{document}